\documentclass[final]{aipproc}

\layoutstyle{6x9}

\begin{document}

\title{Burst Detector Sensitivity: Past, Present \& Future}


\author{David L. Band}{
  address={Code 661, NASA/Goddard Space Flight Center, Greenbelt, MD
  20771},
  altaddress={JCA, University of Maryland, Baltimore County, Baltimore, MD 21250}
}

\begin{abstract}
I compare the burst detection sensitivity of {\it CGRO}'s BATSE,
{\it Swift}'s BAT, the {\it GLAST} Burst Monitor (GBM) and {\it
EXIST} as a function of a burst's spectrum and duration.  A
detector's overall burst sensitivity depends on its energy
sensitivity and set of accumulations times $\Delta t$; these two
factors shape the detected burst population.  For example, relative
to BATSE, the BAT's softer energy band decreases the detection rate
of short, hard bursts, while the BAT's longer accumulation times
increase the detection rate of long, soft bursts. Consequently, {\it
Swift} is detecting long, low fluence bursts (2-3$\times$ fainter
than BATSE).
\end{abstract}

\maketitle

What is the relative sensitivity of different detectors for
detecting gamma-ray bursts, and how should this sensitivity
be compared?  How do these differences shape the observed
burst populations, which must be taken into account in
determining the underlying burst distribution?  Here I
compare BATSE's Large Area Detectors on {\it CGRO} (the
past), the Burst Alert Telescope (BAT)\cite{Barthelmy2005}
on {\it Swift} (the present), and the {\it GLAST} Burst
Monitor (GBM) and {\it EXIST} (the future). BATSE and the
GBM are/were sets of NaI(Tl) detectors while the BAT and
{\it EXIST} are/will be CZT coded mask detectors. The
energy range of NaI(Tl) detectors is $\sim$20--1000~keV
while for CZT it is $\sim$10--150~keV.  I apply a
semi-analytic methodology using simplified models of the
trigger systems of the different detectors.

Most instruments detect bursts using either rate triggers or image
triggers. A rate trigger determines whether the increase in the
number of counts in a time bin $\Delta t$ and energy band $\Delta E$
over the expected number of background counts is statistically
significant. An image trigger determines whether the image formed
from the counts in the time bin $\Delta t$ and energy band $\Delta
E$ contains a new point source. Usually an image trigger is preceded
by a rate trigger that starts the imaging
process\cite{Fenimore2003}; the rate trigger is set to permit many
false positives that are eliminated by the image trigger.  If the
number of burst counts is $S$ and the number of non-burst counts is
$B$, then the rate trigger significance $\sigma_r$ (for BATSE and
the GBM) and the image trigger significance $\sigma_i$ for $\Delta
t$ and $\Delta E$ are
\begin{equation}
\sigma_r = {S\over \sqrt{B}} \qquad \hbox{and} \qquad \sigma_i =
{{f_c S} \over \sqrt{B+S}}
\end{equation}
where $f_c$ accounts for the finite size of the detector pixels. For
{\it Swift} $f_c \sim$0.7\cite{Skinner2005}, which explains why for
a given burst the rate trigger significance is greater than the
imaging significance\cite{Palmer2005}. The BAT uses a more complex
rate trigger than shown above. For directions other than the burst
position the counts $S$ from the burst contribute to the average
flux level, and therefore in imaging $S$ is compared to
$\sqrt{B+S}$. A trigger occurs when $\sigma_r$ or $\sigma_i$ exceeds
a threshold value that is sufficiently high for a small probability
of false positives. Because of the similarity of $\sigma_r$ and
$\sigma_i$, particularly since usually $B\gg S$ near threshold, the
methodologies for evaluating the sensitivity of rate and image
triggers are essentially the same\cite{Band2003}.

Whether a given detector and its trigger system detects a burst
depends on the number of counts $S$ the burst produces in time bin
$\Delta t$ and energy band $\Delta E$. If bursts differed only in
their intensity, we could use a common measure of burst intensity,
but bursts differ in their temporal and spectral properties, and
thus a given detector is more sensitive to some burst types and less
to others. While it is impossible to characterize a burst completely
by only a few parameters, approximate burst types can be described
by a few parameters.  For burst detection, I characterize bursts by:
$E_p$---the energy at the peak of $E^2 N(E)\propto \nu f_\nu$ for
the spectrum averaged over $\Delta t$; and $T_{90}$---the time
duration for 90\% of the flux.  I characterize the burst intensity
by the peak flux $F_T$ over 1~s in the 1--1000~keV band.  A detector
does not measure $F_T$ directly; since a spectral fit is necessary
to convert from counts to flux, $F_T$ need not be over $\Delta
E$\cite{Band2003}.

I first evaluate the sensitivity of the four detectors to $E_p$,
disregarding the burst duration.  Thus I assume that bursts have
constant emission over $\Delta t$=1 s.  Then $S$ is proportional to
the burst spectrum times the detector efficiency integrated over
$\Delta E$. BATSE used only one value of $\Delta E$ at any given
time, typically $\Delta E$=50--300~keV, but later detectors use a
set of $\Delta E$ simultaneously. Figure 1 shows the threshold $F_T$
as a function of $E_p$ for BATSE, BAT, GBM and {\it EXIST}. While I
characterize the spectrum primarily by $E_p$, there remains a
dependence on the high and low energy spectral indices, $\beta$ and
$\alpha$ (not shown by Figure~1). In all cases I use the maximum
sensitivity over the FOV. The scalloping results from multiple
values of $\Delta E$.

\begin{figure}
  \includegraphics[height=.29\textheight]{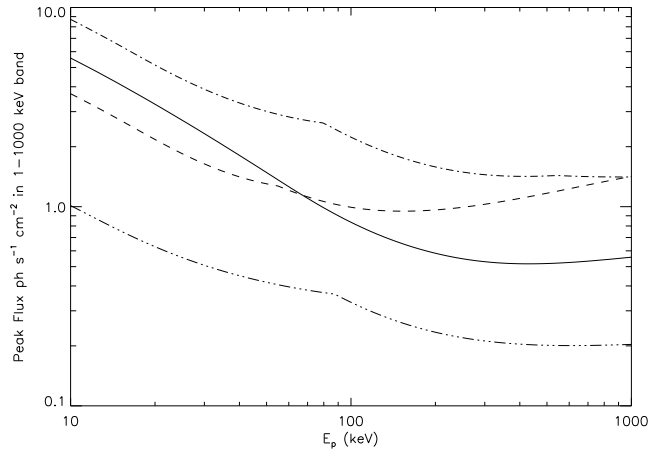}
  \caption{
Threshold value of $F_T$ (1--1000~keV peak flux) as a
function of $E_p$ for BATSE (solid), the BAT (dashed), the
GBM (dot-dashed) and one {\it EXIST} telescope (3
dots-dashed). The spectrum has a low energy spectral index
of $\alpha$= -0.5 and a high energy index of $\beta$= -2. }
\end{figure}

As a CZT detector, the BAT's sensitivity is shifted to
lower energy than BATSE's. The GBM's detectors will be
smaller than BATSE's, while {\it EXIST} will have larger
detectors than the BAT. Note that these sensitivity curves
are all for $\Delta t$=1~s; increasing $\Delta t$ increases
the BAT's sensitivity to long bursts.

\begin{figure}
  \includegraphics[height=.29\textheight]{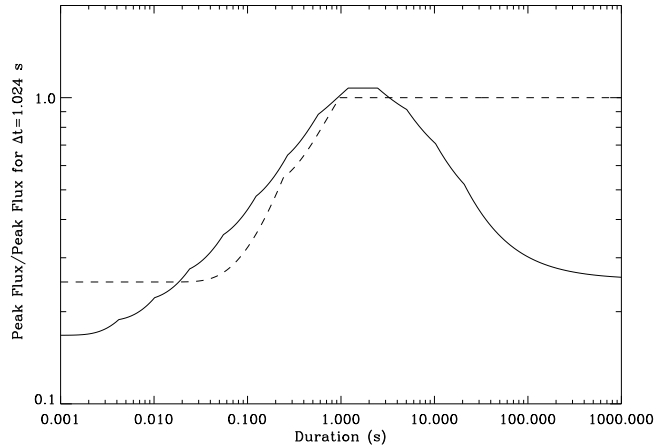}
\caption{The ratio of the detector sensitivity for a
trigger system using a set of $\Delta t$ values to the rate
trigger sensitivity for $\Delta t$=1.024~s alone; ratios
less than 1 indicate an increase in sensitivity resulting
from additional values of $\Delta t$.  The dashed curve is
for BATSE's set of $\Delta t$, while the dashed curve is
for the BAT.}
\end{figure}

As $\Delta t$ increases, the number of burst counts $S$ may increase
but the number of background counts $B$ definitely increases.  Thus
there is a competition between changes in $S$ and $B$ as $\Delta t$
changes. The affect on the sensitivity of increasing or decreasing
$\Delta t$ depends on the burst lightcurve; the duration $T_{90}$ is
a key parameter characterizing the lightcurve.  If $T_{90}$<<$\Delta
t$ then the dilution of the burst counts by the background can be
decreased by decreasing $\Delta t$, but if $T_{90}$>>$\Delta t$ then
increasing $\Delta t$ might increase the number of burst counts
relative to the background. Detectors use more than one $\Delta t$;
the overall sensitivity is the lowest value of $F_T$ for any $\Delta
t$. Figure 2 shows as a function of $T_{90}$ the ratio of the
detector sensitivity for a set of $\Delta t$ to the rate trigger
sensitivity for $\Delta t$=1.024~s alone; ratios less than 1
indicate an increase in sensitivity resulting from the additional
values of $\Delta t$ and the trigger type. The burst is assumed to
have an exponential lightcurve. BATSE (dashed curve) used a simple
rate trigger with $\Delta t$=0.064, 0.256 and 1.024~s. Thus, for
$T_{90}>$1.024 s the $\Delta t$=1.024~s trigger dominates, and the
ratio equals 1, while for $T_{90}<$1.024~s the smaller $\Delta t$
values increase the sensitivity (smaller ratios) to short duration
bursts. The BAT, GBM and {\it EXIST} (will) use both smaller and
longer $\Delta t$ values than BATSE did.  The solid curve on Figure
2 was calculated for the BAT's image trigger; note the increase in
sensitivity over BATSE's set of $\Delta t$ for both long and short
bursts.  The increase in sensitivity to short duration bursts is not
as dramatic as for long duration bursts because $\sigma_i$ does not
decrease indefinitely as $B$ decreases for fixed $S$ (see eq.~1).
The reduced sensitivity of an image trigger relative to a simple
rate trigger for very short bursts does NOT mean that a rate trigger
is superior to an image trigger: an image trigger localizes the
burst, and does not require a model of the background rate.

The spectral and temporal dependencies of the burst sensitivity can
be combined to produce the threshold $F_T$ as a function of both
$E_p$ and $T_{90}$.  Figure 3 presents the ratio of the
sensitivities of the BAT and BATSE; a ratio less than one indicates
that the BAT is more sensitive than BATSE at that particular set of
$E_p$ and $T_{90}$.  Also plotted are the $E_p$ and $T_{90}$ for a
set of BATSE bursts.  As can be seen, the short, hard bursts are in
a region of parameter space where the BAT is less sensitive than
BATSE while the BAT is more sensitive to long, soft bursts. The
contours' gradient shows that the BAT detects fewer short, hard
burst because its energy band is lower than BATSE's was, and the BAT
detects more long, soft bursts both because of its lower energy band
and its greater sensitivity to long bursts. This explains the shift
in the duration distributions. Because the BAT is significantly more
sensitive to long bursts, the average fluence detected by the BAT is
$\sim2.5$ times fainter than BATSE's. As a burst's redshift is
increased, its duration is dilated and its spectrum is redshifted.
Thus high redshift bursts are shifted towards the parameter space
region where {\it Swift} is particularly sensitive. Note that {\it
Beppo-SAX} and {\it HETE-II} use(d) imaging triggers with low energy
detectors, explaining why these two detectors detect(ed) few short
bursts and many X-ray rich bursts and X-ray Flashes.

\begin{figure}
  \includegraphics[height=.29\textheight]{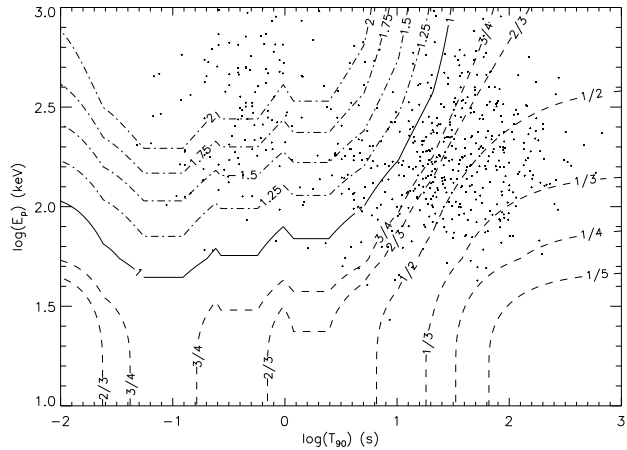}
  \caption{
Contour plot of the ratio of the sensitivities of the BAT and BATSE
as a function of $E_p$ and $T_{90}$; a ratio less than one indicates
that the BAT is more sensitive than BATSE. $\alpha$=-0.5 and
$\beta$=-2 are assumed.  Also plotted are the $E_p$ and $T_{90}$ for
a set of BATSE bursts with enough counts for spectral fits. }
\end{figure}

\begin{theacknowledgments}
I thank members of the BAT instrument team for informative
discussions; conclusions regarding the BAT's sensitivity are mine,
and do not represent the BAT team.
\end{theacknowledgments}

\end{document}